\documentclass[fleqn,a4paper]{vch-book}
\usepackage{amsmath}
\usepackage{amssymb}
\usepackage{makeidx}
    \makeindex
\usepackage{times}
\usepackage{tabularx}
\usepackage{booktabs}
\usepackage[dvips]{epsfig}
\DeclareGraphicsExtensions{.eps,.ps}

\chardef\bslash=`\\ 

\newcommand{\bibtex}{\ifx\is@itshape\f@shape{\fontshape{scit}\selectfont
Bib}\else\textsc{Bib}\fi\kern-.1em\TeX}

\usepackage{bm}

\begin{document}

\chapter{EMS Measurement of the Valence Spectral Function of Silicon - a test of Many-body Theory
}\label{chap:EMS}

\authorafterheading{C.~Bowles, A.S. Kheifets, V.A. Sashin, M. Vos,  E.~Weigold\footnote{e-mail: erich.weigold@anu.edu.au}}

\affil{Atomic and Molecular Physics Laboratories, Research School of Physical
Sciences\\ and Engineering, Australian National University, Canberra, 0200
Australia} \vspace{1.5cm}
\authorafterheading{F.~Aryasetiawan}
\affil{Research Institute for Computational Sciences, AIST, Tsukuba Central 2,
Umezono 1-1-1, Tsukuba Ibaraki 305-8568 Japan }

\section{Introduction}
The electronic properties of the ground states of semiconductors have been studied
both experimentally and theoretically for many years. Thus angle resolved
photoelectron spectroscopy (ARPES), especially in combination with tunable
synchrotron light sources, has been extensively used to map the dispersion of bands
in single crystals. However, in the past the experimental work has concentrated
almost exclusively on the measurement of energies, and the theoretical valence-band
structure calculations have been tested essentially only in terms of their
predictions of eigenvalues. In contrast, relatively little attention has been given
to the wave function of the electrons, despite the fact that wave function
information provides a much more sensitive way of testing the theoretical model
under investigation. Although the wave function cannot be measured directly it is
closely related to the spectral momentum density. In the independent particle model
it is simply proportional to the modulus square of the one-electron wave function,
\begin{equation}
A (\bm{q},\omega) =  \sum_{\bm{G}\bm{k}}n_{j,\bm{k}} |\Phi_{j,\bm{q}}(\bm{q})|^2
\delta(\bm{q}-\bm{k}-\bm{G}) \delta(\omega-\varepsilon_{j\bm{k}}) ,
\end{equation}
where $\Phi (\bm{q})$ is the momentum-space one-electron wave function,  $j$ is the
band index, $\bm{k}$ the crystal wave vector, and $n_{j\bm{k}}$ and
$\varepsilon_{j\bm{k}}$ are the occupation number and energy of the corresponding
one-electron state. The reciprocal lattice vector $\bm G$ translates the momentum
$\bm q$ to the first Brillouin zone. For an interacting many-electron system the
full spectral electron momentum density (SEMD) is given by
\begin{equation}
A (\bm{q},\omega) = \frac{1}{\pi}  G^-(\bm{q},\omega) =\frac{1}{\pi}\frac{1}{[\omega
- h - \Sigma(\bm{q},\omega)]} \label{spec_f}
\end{equation}

Here $G^- (\bm{q},\omega)$ is the interacting single-hole (retarded) Green's
function of the many-electron system, $\Sigma$ is the self energy and $h$ is the
one-electron operator, which includes the kinetic energy and the Coulomb potential
from the nuclei and the average of the electron charge cloud density (the Hartree
potential). Presuming that the Green's function can be diagonalized on an
appropriate basis of momentum-space quasiparticle states $\phi_j(\bm{q})$ (e.g.
orbitals in atoms, Bloch waves in crystals) then for a crystal it takes the form
\cite{Vos02D}
\begin{equation}
A (\bm{q},\omega)  = \sum_{j,\bm{k},\bm{G}} |\phi_j (\bm{q})|^2
\delta_{\bm{q},(\bm{k}+\bm{G})} \frac{1}{\pi} Im G_j^-(\bm{k},\omega) .
\label{green}
\end{equation}
 In the absence of electron-electron interactions the non-interacting Green's
function is simply a delta function and eq. (\ref{green}) reduces to eq.
(\ref{spec_f}). The interacting SEMD contains much more information than simply the
band dispersion. The main feature describes the probability of quasiparticle in band
$j$ having momentum $\bm k$ and energy $\omega$. The center of the quasiparticle
peak is shifted with respect to the one-electron energy $\varepsilon_{j\bm{k}}$ and
it acquires a width due to the finite quasiparticle lifetime. In addition electron
correlation effects can give rise to significant satellite structures.

The full SEMD can be measured by electron momentum spectroscopy (EMS)
\cite{Vos02D,Weigold99}, in which the energies $E_0$, $E_1$ and $E_2$ and momenta
$\bm{k}_0$, $ \bm{k}_1$ and $ \bm{k}_2$ of the incident (subscript 0) and two
outgoing electrons (subscripts $1$ and $2$) in high-energy high-momentum-transfer
$(e,2e)$ ionizing collisions are fully determined. From energy and momentum
conservation one can determine for each $(e,2e)$ event the binding (or separation)
energy of the ejected electron
\begin{equation}
                        \omega = E_0 - E_1 - E_2,
                        \label{encons}
\end{equation}
and the recoil momentum of the ionized specimen
\begin{equation}
                        \bm{q} = \bm{k}_1 + \bm{k}_2 - \bm{k}_0.
                        \label{momcons}
\end{equation}
The differential cross section is given by \cite{Weigold99}.
\begin{equation}
                        \sigma(\bm{k}_0, \bm{k}_1, \bm{k}_2, \omega) =
                        (2\pi)^4 k_0^{-1} k_1 k_2 f_{ee} A(\bm{q},\omega).
\end{equation}
Here $f_{ee}$ is the electron-electron scattering factor, which is constant in the
non-coplanar symmetric high-energy $(e,2e)$ kinematics used in the spectrometer at
the Australian National University \cite{Vos00,Vos00b}. Thus the $(e,2e)$ cross
section is directly proportional to the full interacting SEMD. Since the EMS
measurements involve real momenta, the crystal momentum $\bm k$ not appearing in the
expression for the cross section, EMS can measure SEMDs for amorphous and
polycrystalline materials as well as for single crystals.

The prototype semiconductor silicon has been used as a test-bed to investigate the
influence of electron correlations on the SEMD, $A(\bm{q},\omega)$. Many
first-principles calculations have been carried out on bulk silicon, see e.g. refs
\cite{Fleszar97,Kheifets95,Borrmann87,Sturm92,Hybertsen86,Godby88,Rohlfing93,Engel96}.
The majority of these calculations are based on the \emph{GW} approximation to the
interacting Green's function \cite{Hedin99,Aryasetiawan98}. The dispersion of the
bulk bands in silicon has been studied with ARPES along high symmetry directions
(see e.g. refs \cite{Johansson90,Wachs85,Rich89b}). However, there has been
essentially no experimental data available on the shapes of the quasiparticle peaks
(i.e. quasiparticle lifetimes) as a function of momentum, and the satellite density
as a function of energy and momentum. These properties of the SEMD arise directly
from electron correlation effects and provide stringent tests for approximations to
the many-electron problem. First-principles calculations of many physical quantities
of interest require the interacting one-particle Green's function as input. It is
therefore important to have reliable methods for accurately calculating and testing
the real and imaginary parts of the self energy, and hence of $A(\bm{q},\omega)$.

There are severe difficulties in extracting the full $A(\bm{q},\omega)$ from
experimental data obtained by other techniques. In ARPES these difficulties include
knowing the specifics of the transitions involved, such as the untangling of
final-state effects from the initial-state ones, and the strong energy and momentum
dependence of the matrix elements \cite{Hedin99,Bansil02}. ARPES is also very
surface sensitive, which can obscure details of the bulk electronic structure (see
e.g. \cite{Rich89b}). In addition there is usually a significant background
underlying the photo-electron spectrum and this can hide any continuous satellite
contributions. Certain Compton scattering experiments in which the struck electron
is detected in coincidence with the scattered photon, so-called $(\gamma ,e\gamma)$
experiments, can in principle map out the spectral function \cite{Bell01,Sattler01}.
However, the energy resolution is such that it is extremely difficult to resolve
even the valence contributions from that of the core electrons \cite{Itou99}.

In this paper we present EMS measurements of the valence spectral momentum density
for the prototypical and most studied semiconductor, silicon, and compare the
results with calculations based on the independent particle approximation as well as
calculations based on many-body approximations to the interacting one-particle
Green's function. In section \ref{secII} we discuss the experimental technique. The
theoretical models are outlined in section \ref{secIII}. The results are discussed
and compared with the  FP-LMTO calculations and the first-principles many-body
calculations in section \ref{secIV} and, where appropriate, with previous ARPES
data. In section \ref{secIV} the role of diffraction is also discussed and in the
last section a brief summary and conclusion is given.
\begin{vchfigure}
\includegraphics[width=0.7\textwidth,keepaspectratio=true,clip]{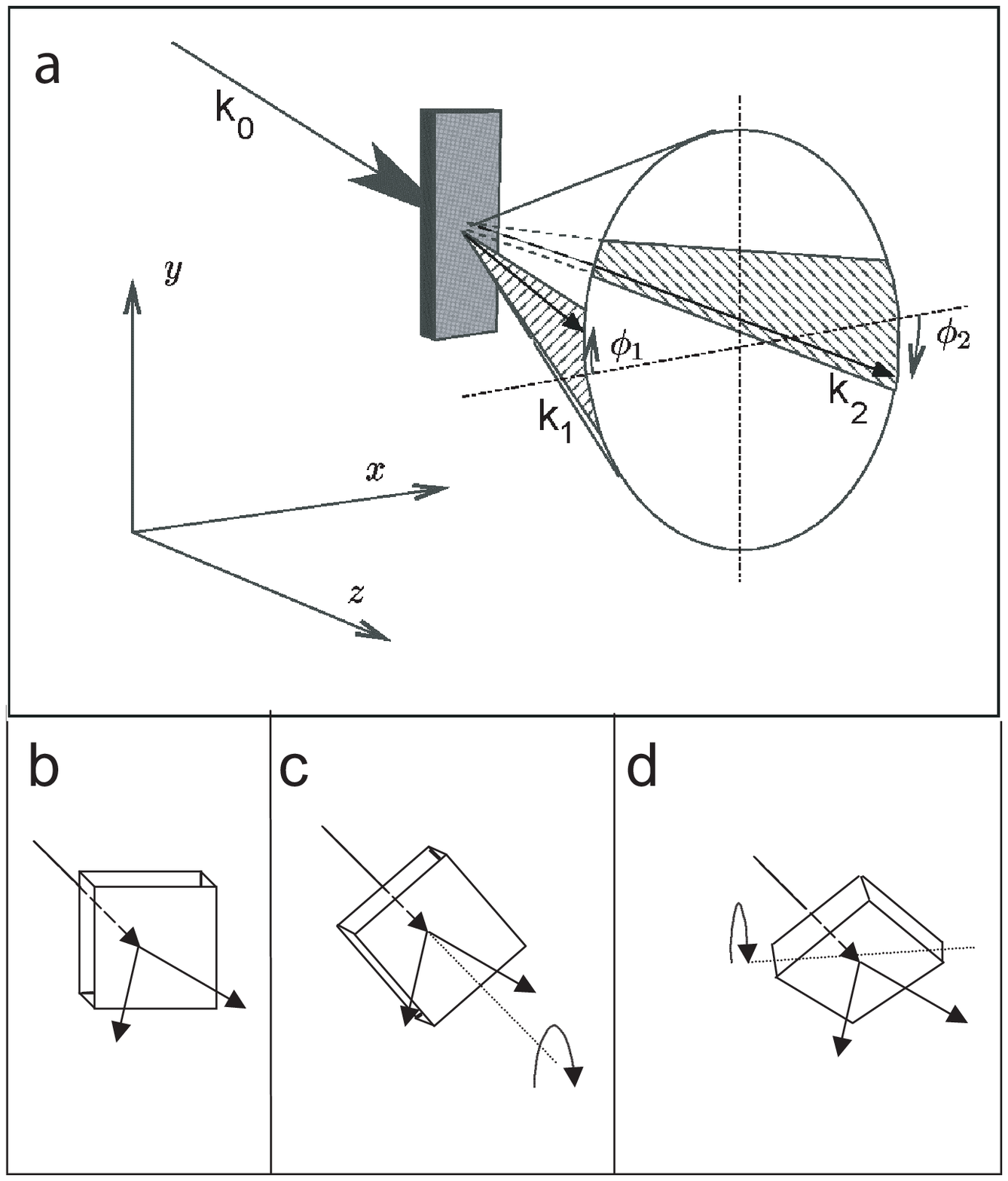}
\vchcaption{Schematics of the experimental arrangements. Incident electrons of
momentum $\bm{k}_0$ along the $z-$axis eject an electron from a thin self-supporting
Si crystal. The scattered and ejected electrons emerging along the shaded portions
of the cone defined by $\Theta = 44.3^\circ$, are detected in coincidence by two
energy and angle sensitive analysers.  In the bottom panels the sample is indicated
as a block with sides parallel to the $\langle 0 1 0 \rangle$ and $\langle 0 0 1
\rangle$ symmetry directions. Thus in (b) the spectral momentum density is measured
along the $\langle 0 1 0 \rangle$ direction.  In (c) the density is  measured along
the $\langle 1 1 0 \rangle$ direction, as the crystal has been rotated about the
surface normal by $45^\circ$. In (d) the crystal has been tilted by $35.3^\circ$
relative to the position in (c)   so that the density is measured along the $\langle
1 1 1 \rangle$ direction.} \label{emsfig1}
\end{vchfigure}

\section{Experimental Details} \label{secII}
An outline of the experimental apparatus (described fully in \cite{Vos00,Vos00b})
and the coordinate system is shown in Fig. \ref{emsfig1}. An electron gun emits a
highly collimated 25 keV electron beam, which enters the sample region inside a
hemisphere held at +25 kV. Thus 50 keV electrons impinge on the target sample along
the $z-$direction, the diameter of the beam being 0.1 mm. The emerging pairs of
electrons with energies near 25 keV are decelerated and focussed at the entrance of
two symmetrically mounted hemispherical electrostatic analyzers. The analyzers
detect electrons emerging along sections of a cone defined by $\Theta_s =
44.3^\circ$, which is chosen so that if all three electrons are in the same plane
then there is no momentum transferred to the target (i.e. $q = 0$). In the
independent particle picture this corresponds to scattering from a stationary
electron. If the electrons are not in the same plane (i.e. $\phi_1 \ne \phi_2$) then
there is a $y-$component of momentum with $q_x = q_z = 0$, that is only target
electrons with momentum ${\bm q}$ directed along the $y-$axis can cause a
coincidence event. The electrons are detected by two-dimensional position-sensitive
electron detectors, mounted at the exit planes of the analyzers, which measure
simultaneously over a range of energies and $y-$momenta (i.e. range of angles
$\phi$).

Two pairs of deflection plates mounted inside the high-voltage hemisphere along the
sections of the cone can be used to change the effective scattering angle by up to
$1^\circ$. In this way \cite{Vos00} one can select nonzero values for the $x-$
and/or the $z-$component of momentum, the $y-$component always lying in the range
0-5 a.u. (atomic units are used here, 1 a.u. = 1.89~\AA$^{-1}$). This allows one to
probe the full three-dimensional momentum space. The two double deflectors are also
used to check that the measured momenta correspond to the expected ones
\cite{Vos00,Vos04}, ensuring that there are no offsets in $q_x$ or $q_z$ as a result
of any small possible geometrical misalignments. The orientation of the target
specimen can be determined by observing the diffraction pattern of the transmitted
electron beam on a phosphorus screen. A specific direction of the thin crystal
target is then aligned with the $y-$axis of the spectrometer so that we measure the
energy-resolved momentum density along that direction. The sample orientation can be
changed without removing the sample from the vacuum by means of the manipulator
mounting arrangement \cite{Vos00}.

The single crystal target has to be an extremely thin self-supporting film. The
initial part of the target preparation followed the procedure of Utteridge et al
\cite{Utteridge00}. First a buried silicon oxide layer was produced by ion
implantation in a crystal with $\langle 1 0 0 \rangle$  surface normal. A crater was
then formed on the back of the crystal by wet chemical etching. The oxide layer
serves to stop the etching. It is then removed by a HF dip and the sample is
transferred to the vacuum. At this stage the thickness of the thinned part of the
crystal is around 200 nm. Low energy (600 eV) argon sputtering is then used to
further thin the sample. The thin part of the crystal is completely transparent. The
thickness is monitored by observing the colour of the transmitted light from an
incandescent lamp placed behind the sample. The colour changes with thickness due to
the interference of the directly transmitted light with that reflected from the
front and back silicon-vacuum interfaces. The thinning is stopped when the thickness
reaches 20 nm (corresponding to a grey-greenish light). The sample is then
transferred to the spectrometer under UHV. A thin amorphous layer could be present
on the backside due to the sputtering, whereas the front side (facing the analyzers)
is probably hydrogen terminated as a result of the HF dip. The base pressure in the
sputtering chamber was of the order of $10^{-9}$ Torr and in the spectrometer the
operating pressure was $2\times 10^{-10}$ Torr.

The experimental energy and momentum resolutions were discussed in detail by Vos et
al \cite{Vos02D}. The full width at half-maximum (FWHM) energy resolution is 1.0 eV,
whereas the FWHM momentum resolutions are estimated to be (0.12, 0.10, 0.10 a.u.)
for the $q_x, q_y, q_z$ momentum components respectively.

\begin{vchfigure}
\includegraphics[width=0.5\textwidth,keepaspectratio=true,clip]
{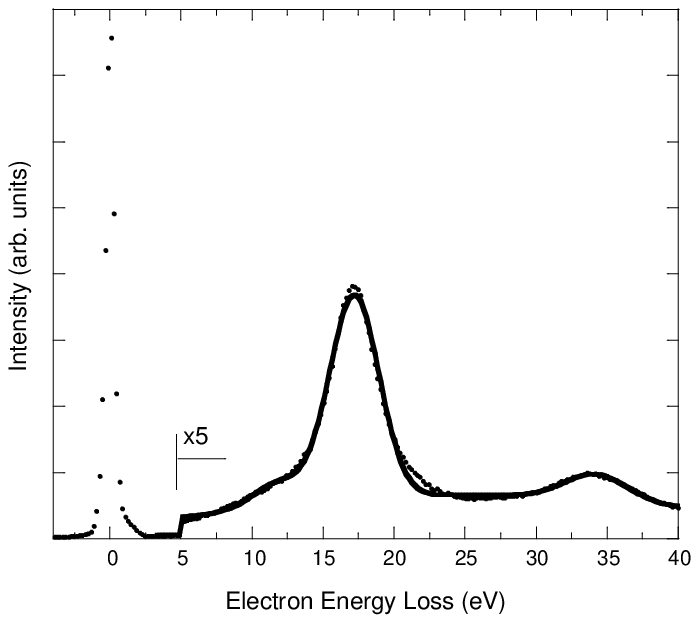} \vchcaption{The energy loss spectrum observed in the spectrometer for
the 20 nm thick Si crystal sample with 25 keV incident electrons. The fitted curve
is used to deconvolute the data for inelastic scattering.} \label{emsfig2}
\end{vchfigure}
Even for a very thin target multiple scattering by the incoming or outgoing
electrons has to be taken into account. These fast electrons can loose energy by
inelastic collisions, such as plasmon excitation, or transfer momentum by elastic
collisions (deflection from the nuclei). These scattering events move real
coincident $(e,2e)$ events to "wrong" parts of the spectrum, as either the energy or
momentum conservation equations (eqs. \ref{encons} and \ref{momcons}) are used
incorrectly. For polycrystalline or amorphous solids these multiple-scattering
events can be modeled by Monte Carlo simulations \cite{Vos96}. In the case of
inelastic scattering the binding energy as inferred from eq. \ref{encons} will be
too high. Inelastic multiple scattering events can be deconvoluted from the data by
measuring an energy loss spectrum for 25 keV electrons passing through the sample.
This deconvolution is done without using any free parameters \cite{Vos02c}. This
approach is used in the present case. Figure \ref{emsfig2} shows the energy loss
spectrum obtained with the present silicon crystal target. For single crystals
elastic scattering from the nuclei adds up coherently (diffraction). The change of
the incoming or outgoing momenta by diffraction changes the outcome of the
measurement by a reciprocal lattice vector. We demonstrate later in this paper how
different measurements can be used to disentangle the diffracted contributions from
the primary (non-diffracted) contribution.

\section{Theory} \label{secIII}

\subsection{Independent Particle Approximation}

The local density approximation (LDA) of density functional theory (DFT) has long
been established as a very useful tool for investigating ground-state properties of
bulk semiconductors from first principles \cite{Jones89,Dreizler90}. The advantage
of DFT for approximate calculations in many-body systems is that one extracts the
needed information from a one-body quantity, the electron density
$n({\boldsymbol{r}})$. Although the one-particle eigenvalues in the theory have no
formal justification as quasiparticle energies, in practice they turn out to be
surprisingly accurate \cite{Pickett89}.

We employed here the linear-muffin-tin-orbital (LMTO) method \cite{Skriver84} within
the framework of DFT. The LMTO method is just one of many computational schemes
derived within the framework of the DFT. The great practical advantage of the LMTO
method is that only a minimal basis set of energy-independent orbitals (typically
9-16 per atom) is needed to obtain accurate eigenvalues (band energies).  In the
present study we implemented a full-potential version (FP-LMTO) of the method
\cite{Kheifets99}. We write the one-electron wave function in a crystal in the
tight-binding representation as the Bloch sum of the localised MT orbitals:
\begin{equation}
\Psi_{j k} (\bm{r}) = \sum_{t} e^{i{\bm k}\cdot {\bm t}} \sum_{\Lambda} a^{j
\bm{k}}_\Lambda \phi_\Lambda ({\boldsymbol {r }} - {\boldsymbol {R}} - \bm{t}) \; .
\label{e17}
\end{equation}
Here $\bm k$ is the crystal momentum, $j$ band index, $\bm t$ lattice (translation)
vector and $\bm R$ basis vector. The label $\Lambda$ defines a MT orbital centered
at a given site $\bm R$ and it comprises the site index $\bm R$ and a set of
atomic-like quantum numbers. The expansion coefficients $a^{j \bm{k}}_{\Lambda}$ are
found by solving the eigenvalue problem using the standard variational technique.

Momentum space representation of the wave function $\Psi_{j\bm{k}}$ is given by the
Fourier transform of the Bloch function:
\begin{equation}
\label{bloch} \Phi_{j{\bm k}}(\bm{q})= \int e^{-i{\bm q}\cdot {\bm
r}}\Psi_{j\bm{k}}({\bm r}) d{\bm r}.
\end{equation}
Due to the periodic nature of the charge density the only non-zero contributions to
$\Phi_{j\bm{k}}(\bm{q})$ occur when ${\bm q}= {\bm k} + {\bm G}$ with $\bm G$ a
reciprocal lattice vector.
\begin{equation}
\Phi_{j\bm{k}} (\bm{q}) = \sum_{\bm G} c_{{\bm G},{\bm k}}^j \delta ({\bm q} - {\bm
k} - {\bm G}). \label{Bloch_comp}
\end{equation}

 The contributions $c^j_{{\bm G},{\bm k}}$, the Bloch wave amplitudes, are expressed through
the (Fourier) integrals:
\begin{eqnarray}
\label{fourier} c^j_{{\bm G},{\bm k}}&=&\Omega^{-1} \int
e^{-i(\bm{k}+\bm{G})\cdot\bm{r}} \Psi_{j\bm{k}}(\bm{r})d\bm{r}
\\ & = &
\Omega^{-1} \sum_\Lambda [ a^{j{\bf k}}_\Lambda e^{-i(\bm{k}+\bm{G})\cdot\bm{R}}
\int e^{-i(\bm{k}+\bm{G})\cdot\bm{r}} \phi_\Lambda(\bm{r}) d\bm{r} ]\nonumber
\end{eqnarray}
Here it is assumed that the wave function $\Psi_{j\bm{k}}$ is normalized in the unit
cell of the volume $\Omega$. The limits of the three-dimensional integration
indicates symbolically the whole coordinate space. The EMD in the occupied part of
the band $j$ is proportional to the modulus squared Bloch amplitudes:
\begin{equation}
\label{rho} \rho_j(\bm{q})= {\Omega^2\over (2\pi)^{3}} \sum_{\bm{G}} n_{j\bm{k}}
\left| \Phi_j(\bm{k}+\bm{G})\right|^2\delta_{\bm{q},\bm{k}+\bm{G}}
\end{equation}
where $n_{j\bm{k}}$ is the occupation number. The EMD (\ref{rho}) is normalized to
the total number of valence electrons per unit cell:
\begin{equation}
\label{sum}
2 \sum_j                                           
\int  \,d\bm{q} \,                       
\rho_j(\bm{q})=                          
n_e\ .
\end{equation}

\subsection{Electron Correlation Models}\label{el_cor}

The hole Green's function entering equation \ref{green} can be calculated by the
many-body perturbation theory (MBPT) expansion on the Bloch wave basis (\ref{e17}).
Taking the first non-vanishing term in the MBPT leads to the so-called \emph{GW}
approximation \cite{Hedin65,Hedin69}. In this acronym \emph{G} stands for the
Green's function and \emph{W} denotes the screened Coulomb interaction. The
\emph{GW} approximation is known to give accurate quasiparticle energies
\cite{Aryasetiawan98}. However, its description of satellite structures is not
satisfactory. In alkali metals, for example, photoemission spectra show the presence
of multiple plasmon satellites whereas the \emph{GW} approximation yields only one
at too large an energy. This shortcoming of the \emph{GW} approximation has been
resolved by introducing vertex corrections in the form of the cumulant expansion to
the Green's function \cite{Langreth70,Bergersen73,Hedin80}. This allowed the
inclusion of multiple plasmon creation. As a result the calculated peak positions of
the plasmon satellites were found to be in much better agreement with the experiment
than those predicted by the \emph{GW} scheme itself
\cite{Aryasetiawan96,Vos99,Vos01}.

Formally, the cumulant expansion for the one-hole Green's function can be derived as
follows. We choose the time representation for the Green's function, drop the band
index $j$ for brevity, and write it as
\begin{equation}
    G(\bm{k}, t < 0) = i \theta (-t) e^{-i\omega_{k}t+C^{h}({\bm k},t)},
    \label{e18}
\end{equation}
where $\omega_{k}$ is the one-electron energy and $C^{h}(\bm{k}, t)$ is defined to
be the cumulant. Expanding the exponential in powers of the cumulant we get
\begin{equation}
    G (\bm{k},t) = G_{0}(\bm{k},t) \left[ 1 + C^{h}
    (\bm{k},t) + \frac{1}{2}
    \left[C^{h}(\bm{k},t)\right]^{2} + \ldots \right],
    \label{e19}
\end{equation}
where $G_{0}(\bm{k},t) = i\theta (-t) \exp (-i\omega_{k}t)$. In terms of the
self-energy $\sum$, the Green function for the hole can be expanded as
\begin{equation}
    G= G_{0}+ G_{0}\sum G_{0} + G_{0}\sum G_{0} \sum G_{0} + \ldots \; .
    \label{e20}
\end{equation}
To lowest order in screened interaction \emph{W}, the cumulant is obtained by
equating
\begin{equation}
    G_{0}C^{h} = G_{0}\sum G_{0},
    \label{e21}
\end{equation}
where $ \sum = \sum_{GW} = i G_{0}W$. The first-order cumulant is therefore
\begin{equation}
    C^{h} (\bm{k},t) = i \int^{\infty}_{t} dt' \int^{\infty}_{t'}
    d\tau e^{i\omega_{k}\tau} \sum (\bm{k},t) \; .
    \label{e22}
\end{equation}
This is then put back into Eq.\ (\ref{e18}) yielding multiple plasmon satellites.
The energy-momentum representation of the Green's function can be restored by the
time Fourier transform.

\begin{vchfigure}
\includegraphics[width=0.9\textwidth,keepaspectratio=true,clip]
{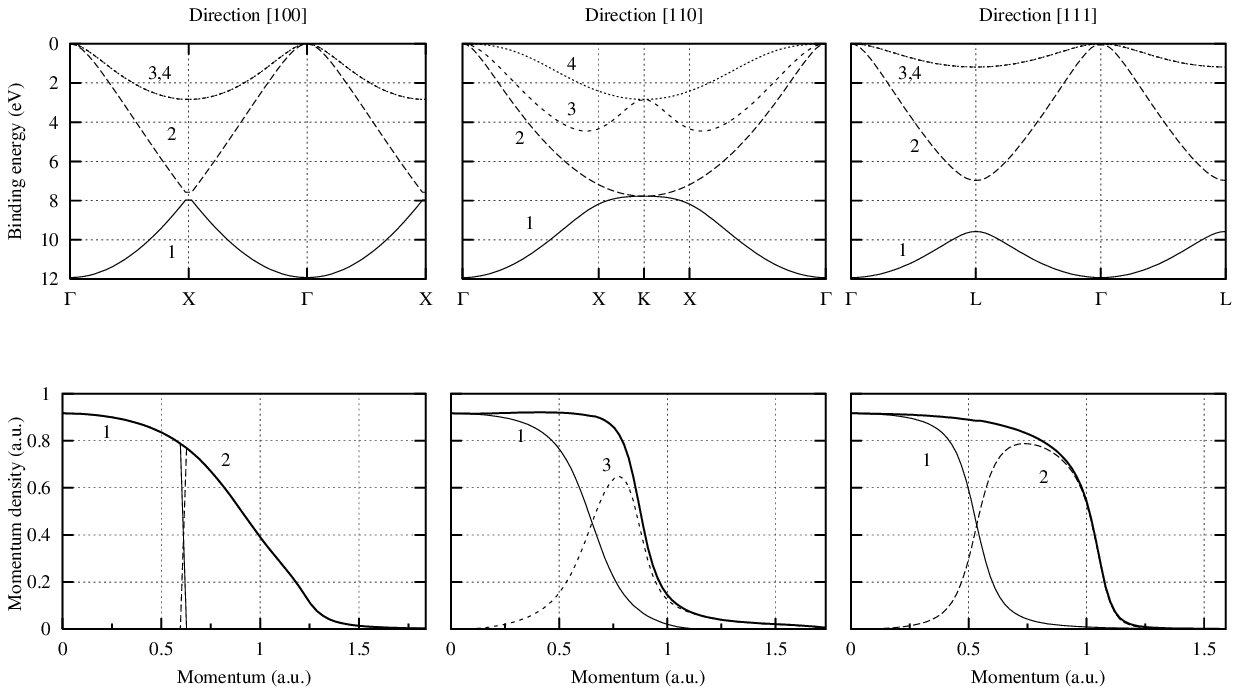} \vchcaption{FP-LMTO calculations \cite{Kheifets99} of the dispersion
(top panel) and momentum density (bottom panel) of Si for different high symmetry
crystallographic directions.  The total momentum density is split up into the
densities of the individual bands as indicated.} \label{emsfig3}
\end{vchfigure}
\section{Results and Discussions} \label{secIV}

\subsection{Band Structure}

We first discuss the dispersion $\varepsilon_{j{\bm k}} (\bm {q})$, i.e. the
dependence of the energy of the Bloch function $\Phi_{j{\bm k}} (\bm{q})$ of band
$j$ on its crystal momentum $\bm{k} = \bm{q} + \bm{G}$.  The band with the largest
binding energy is labeled 1, the next one 2 etc. In momentum space the Bloch
function with crystal momentum $\bm{k}$ is non-zero only at momentum values $\bm{k}
+ \bm{G}$ with amplitude $c^j_{{\bm G},{\bm k}}$ (see eq. \ref{Bloch_comp}).

The band structure and momentum densities obtained by Kheifets et al.
\cite{Kheifets99} in a FP-LMTO calculation (based on the DFT-LDA) are shown in Fig.
\ref{emsfig3}, the bands being labeled as discussed above. The bands are periodic in
${\bm q}$ space, with band 1 having a maximum in the binding energy at the $\Gamma$
points. However, the only $\Gamma$ point with significant momentum density in band 1
is the one corresponding to zero momentum (see lower panel in Fig. \ref{emsfig3}).
Thus the function with the lowest energy is a Bloch function with $k = 0$, $
c^1_{(0,0,0),{\bm k}} \simeq 1$, and the other $c^1_{{\bm G},{\bm k}} \simeq 0$.

The cut of the first four Brillouin zones of silicon along the $q_z = 0$ plane is
shown in Fig. \ref{emsfig4}. For a free electron solid the wave functions are plane
waves and in the ground state the occupied states are within the Fermi sphere with
radius $k_f$ i.e. $|k| < k_f$. The intersection of this sphere with the $q_z = 0$
plane is indicated by the dashed circle in Fig. \ref{emsfig4}. The lattice potential
of silicon can be viewed as a perturbation on the free electron picture, so that the
wave functions are Bloch functions with more than one $|c^j_{\bm{G k}} |^2 > 0$. The
semiconductor silicon has 8 valence electrons per unit cell, and hence the first 4
bands are fully occupied with one spin up and one spin down electron per band. We
will now discuss the results of the measurement of silicon and emphasize that the
electron density for band $j$ is at its maximum in Brillouin zone $j$.

\begin{vchfigure}
\includegraphics[width=0.55\textwidth,keepaspectratio=true,clip]
{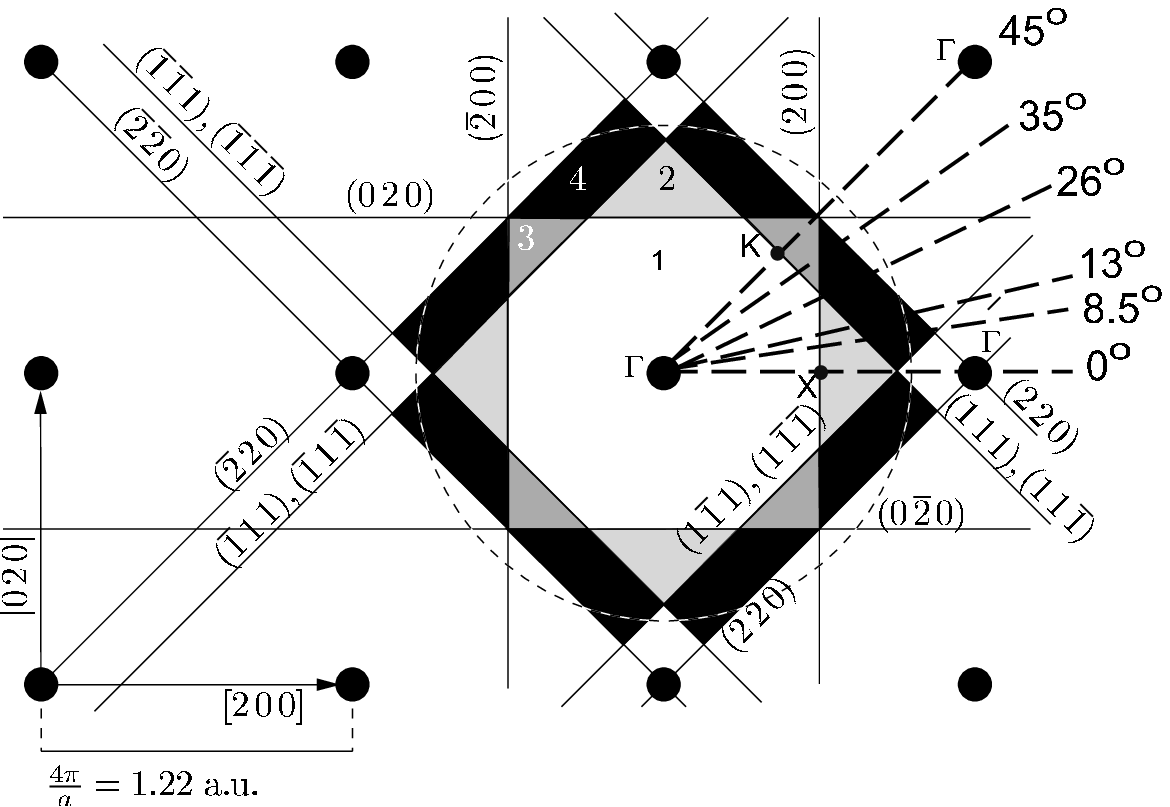} \vchcaption{The cut through the reciprocal lattice of silicon
along the $q_z$ = 0 plane with the first 4 Brillouin zones labeled. The Brillouin
zone boundaries are labeled by the indices of the reciprocal lattice it bisects. The
dashed circle is the Fermi sphere for a free electron solid with the same electron
density as silicon. Different measurements through the $q = 0$ point are indicated
by the dashed lines.} \label{emsfig4}
\end{vchfigure}
The sample was a thin ($\simeq$ 20 nm) single silicon crystal with $\langle 0 0 1
\rangle$  surface normal, which is first aligned with the z direction (i.e. aligned
with ${\bm k}_0$, see Fig. \ref{emsfig1} (a)). Rotating around the surface normal,
measurements were taken with the sample $\langle 1 0 0 \rangle$  direction aligned
along the $y-$axis, then the $\langle 1 1 0 \rangle$ direction and 4 intermediate
directions were aligned along the $y-$axis as shown in Fig. \ref{emsfig4}. In all
these cases the potentials on the sets of deflector plates were set to ensure that
the measurements passed through zero momentum (corresponding to $\Gamma_{(0,0,0)}$).
\begin{vchfigure}
\includegraphics[width=0.85\textwidth,keepaspectratio=true,clip]
{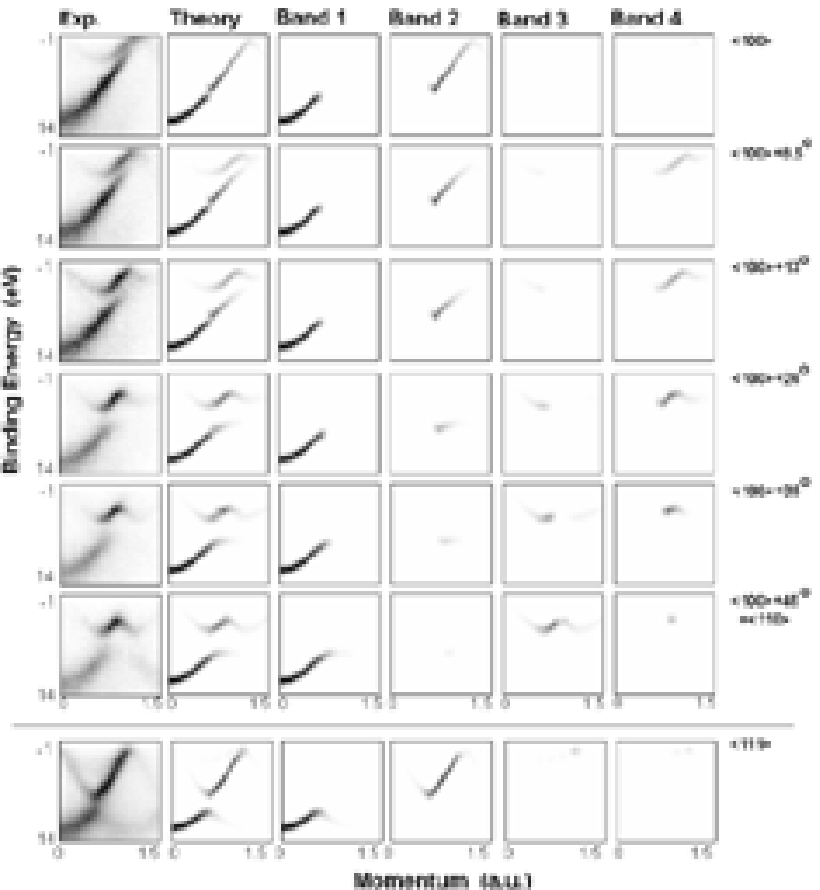}
 \vchcaption{ The LMTO calculated spectral momentum densities (total plus
separate band contributions) along the directions shown in Fig. \ref{emsfig4}
compared with the measured results. The calculations have been broadened by the
experimental energy resolution. The bottom panel shows the results along the
$\langle 1 1 1 \rangle$ direction.} \label{emsfig5}
\end{vchfigure}
The experimentally observed density distributions, together with the results of the
FP-LMTO calculation, are shown in Fig. \ref{emsfig5} for the 6 directions measured
through $\Gamma_{(0,0,0)}$. The calculations were broadened with the experimental 1
eV energy resolution and split up into the 4 occupied bands.

For the measurement of momenta directed along the $\langle 1 0 0 \rangle$ direction
($0^\circ$ in Fig. \ref{emsfig4}) the theory predicts bands 1 and 2 occupied. In the
first Brillouin zone band 1 is occupied, changing abruptly to band 2 at 0.61 a.u.
(see also Fig. \ref{emsfig4} and left panel of Fig. \ref{emsfig3}). There is no band
gap in the dispersion on crossing the first Brillouin zone. This is due to
additional symmetry of the diamond lattice (see e.g. \cite{Heine60}). After leaving
the second Brillouin zone the calculated density drops only gradually to zero. The
measured density (left panel of Fig. \ref{emsfig5}) has the same behaviour , but
also shows an additional branch at smaller binding energies, which merges with the
main feature at 1.2 a.u. This additional branch can also be seen in the calculated
band structure for the case where the crystal has been rotated by $8.5^\circ$, and
comes from band 4. From the shape of the Brillouin zone shown in Fig. \ref{emsfig4}
it is clear that the measurement along the $\langle 1 0 0 \rangle$ direction just
misses Brillouin zone 4. Due to finite momentum resolution it is obvious that the
measurement will pick up contributions from this zone, giving rise to the extra
branch in the observed intensity.

In the $\langle 1 1 0 \rangle$ symmetry direction (reached by a rotation by
$45^\circ$ along the surface normal) the first Brillouin zone crossing is two
planes, (the $(1 1 1)$ and $(1 1 \overline{1})$ planes), making an angle of $\pm
54.35^\circ$ with the $q_z = 0$ plane. Thus the band switches from 1 to 3 at the
double crossing. This gives rise to the classic band gap behaviour, band 1 having a
minimum in binding energy (maximum in energy) at the Brillouin zone crossing, with
its density petering out after the crossing. Band 3 slowly increases in intensity
from zero momentum up to the first Brillouin zone boundary, where it has a maximum
in binding energy, with increased density as one passes through Brillouin zone 3.
The next extremum in energy, which corresponds to a minimum in binding energy, is
when band 3 crosses the next set of Brillouin zone boundaries, i.e. on leaving
Brillouin zone 3, and its intensity decreases thereafter as the momentum increases.
The calculations and measurements are in quite good agreement with each other for
these general features. For the intermediate angles band 1 remains dominant, band 4
is prominent, band 2 makes a significant contribution for directions not far from
the $\langle 1 0 0 \rangle$ direction, and band 3 make small contributions close to
the $\langle 1 1 0 \rangle$ direction.

Also shown in Fig. \ref{emsfig5} (bottom panels) is the spectral momentum density
obtained along the $\langle 1 1 1 \rangle$ direction. This direction was reached by
tilting the $\langle 1 1 0 \rangle$ aligned sample over $35.3^\circ$ (see Fig.
\ref{emsfig1}(d)). Here the density is due to bands 1 and 2 with a large band gap at
the zone crossing. Again these general features of the experiment and theory are in
reasonable agreement. Note that the dispersion in the $\langle 1 1 1 \rangle$
direction could be mapped over a much larger momentum range than expected based  on
the momentum density distribution (see Fig. \ref{emsfig3} bottom panel). This is due
to diffraction, as is described later.
\begin{vchfigure}
\includegraphics[width=0.85\textwidth,keepaspectratio=true,clip]
{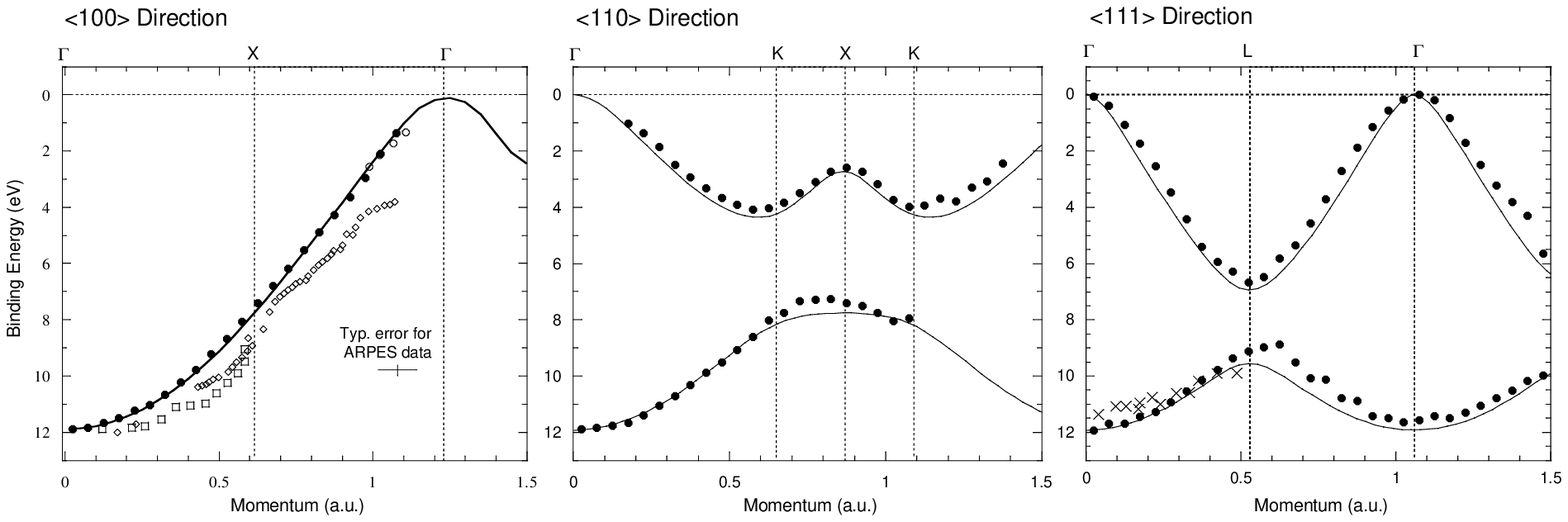} \vchcaption{ The measured dispersion in the peak density (dots) along
the $\langle 1 0 0 \rangle$(left panel), $\langle 1 1 0 \rangle$ (central) and
$\langle 1 1 1 \rangle$ (right panel) symmetry directions. The full line is the LMTO
calculation. Only those branches of the band structure are shown that are expected
to have non-zero intensity in the EMS measurement. The open circles are ARPES data
from ref. \cite{Wachs85} and the diamonds, squares and crosses are ARPES data from
\cite{Rich89b}.} \label{emsfig6}
\end{vchfigure}

A more detailed comparison between the calculated and measured dispersion of the
bands in the three high symmetry directions is shown in Fig. \ref{emsfig6}. Here a
fitting procedure was used to extract the peak position of the measured density at
different momenta. For comparison we include for the $\langle 1 0 0 \rangle$ and
$\langle 1 1 1 \rangle$ directions  also  the ARPES data of refs. \cite{Wachs85} and
\cite{Rich89b}. The agreement between the calculation and the present measurement is
excellent. There is a small deviation near the $X-$point in the $\langle 1 1 0
\rangle$ direction, with the observed gap between bands 1 and 3 being a little
smaller than given by theory. A similar deviations for the band gaps is observed
near the $L$ points  in the $\langle 1 1 1 \rangle$ direction. Generally the EMS
shapes are smoother and better defined than the ARPES ones.

Fig. \ref{emsfig7} shows the band structures measured with the $\langle 1 1 0
\rangle$ direction aligned with the $y-$axis at three different values of $q_x$,
i.e. 0, 0.65 a.u., and 0.87 a.u., the latter two being along Brillouin zone
boundaries. The offsets in the $x-$component of momentum are produced by choosing
suitable settings for the deflector voltages for each set of deflectors
\cite{Vos04}. For the measurement with $q_x = q_z = 0$, there is a single dispersing
feature with a band gap at $q_y = 0.65$ a.u. due to the crossing of the zone
boundary. This momentum value corresponds to a $K$ point of the band structure.
Since the $x-$ and $y-$directions through the $\Gamma$ point are equivalent, the
binding energy spectra at the two $K$ points, $(q_x, q_y, q_z) = (0, 0.65, 0)$ and
$(0.65, 0, 0)$, should be identical. This is indeed the case \cite{Vos04}.
\begin{vchfigure}
\includegraphics[width=0.7\textwidth,keepaspectratio=true,clip]
{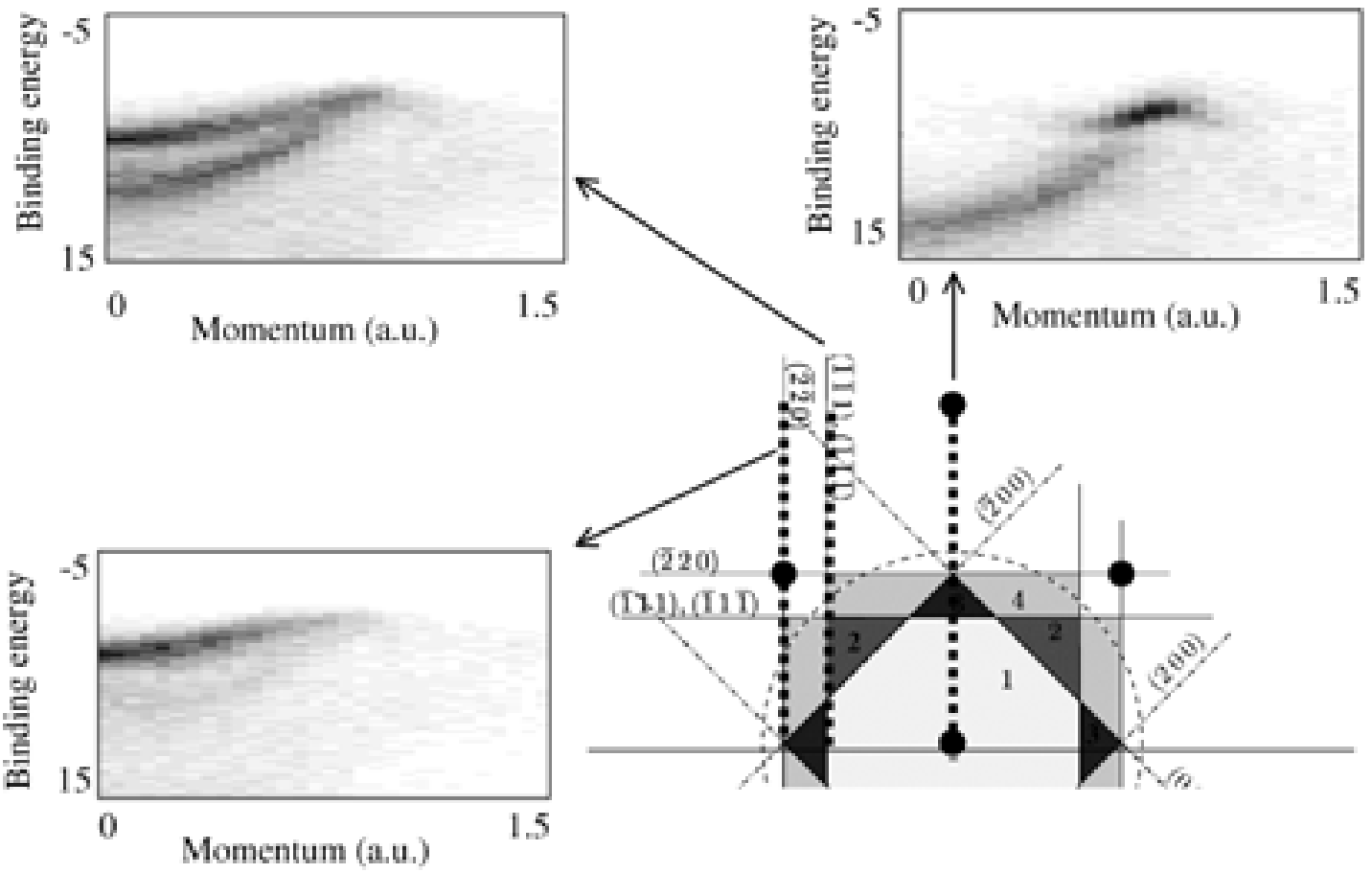} \vchcaption{The measured densities along the lines indicated in the
Brillouin zone cut.  The $x-$axis of the spectrometer coincides with the crystal
$\langle 1 \overline{1} 0 \rangle$ direction. One line goes through the $\Gamma$
point (zero momentum), one has a constant $q_x$ component of $0.65$ a.u. and one has
a constant $q_x$ component of $0.87$ a.u. The latter two lines coincide with
Brillouin zone edges.} \label{emsfig7}
\end{vchfigure}

The total band width can be obtained from the zero momentum spectrum of the
measurement along the $\langle 1 1 1 \rangle$ direction.  Here both the top and the
bottom of the band are clearly visible.  The experimental separation between the two
peaks is 11.85 $\pm$ 0.2 eV.  The error is mainly due to the uncertainty in the
shape of the background under the high-binding energy peak. The LMTO calculation
gives a value of 11.93 eV.

As an alternative way of determining the valence band width we measured the valence
band and Si 2p core level in the same experiment.  This allows us to determine
precisely the position of the valence band bottom relative to the 2p$_{3/2}$
component of the core line.  The fit result gives a value of 86.65 $\pm$0.2 eV for
their energy separation. From photoemission data (e.g.\cite{Himpsel83}) we know that
the 2p$_{3/2}$ binding energy relative to the valence band maximum is 98.74 eV. In
this way we obtain a value of 12.1 $\pm$0.2 eV for the valence band width.

\subsection{Diffraction Effects}

In a crystal coherent elastic scattering by the nuclear sites (dynamic diffraction)
of the incident and/or the emitted electrons can shift the observed momentum
distribution by a reciprocal lattice vector ${\bm G}$
\cite{Vos03B,Allen90,Matthews93}. In the present experiment the electron momenta are
all very large ($k_0 =62.1$ a.u., $k_{1,2} = 43.4$ a.u.), so that for the dominant
diffraction the vectors ${\bm G}_i$ are all much smaller than the respective ${\bm
k}_i$. Thus the ${\bm G}_i$ must be essentially perpendicular to the ${\bm k}_i$
vector to fulfil the diffraction condition $2{\bm k}_i\cdot{\bm G}_i + G_i^2 =0$. In
the case where the surface normal (the $\langle 1 0 0 \rangle$ direction) is aligned
with ${\bm k}_0$ the smallest ${\bm G}_0$ contributing to diffraction are of type
$\langle \pm\! 2 \pm\!\! 2 \;0 \rangle$ (see ref. \cite{Vos03B} for a detailed
discussion). The smallest vector ${\bm q}$ that can be accessed is 1.22 a.u. where
the density is very low (see Fig. \ref{emsfig3}), and it appears at $q_y = 1.22$
a.u. Diffraction events of the type $\langle 0 \pm\!\! 4 \;0 \rangle$ produce an
offset in the $y-$component of momentum by 2.44 a.u. The measured y-momentum line
still passes through the origin so that we can still measure electrons with $q = 0$.
Thus this diffracted beam causes a weak replication of the main spectrum shifted by
2.44 a.u. For the outgoing electrons the diffraction depends on which crystal
direction is aligned with the $y-$axis of the spectrometer \cite{Vos03B}.

In the apparatus it is possible to rotate the sample about the $y-$axis. This does
not affect the direction of measurement, but it does affect the directions of the
incoming and outgoing electrons relative to the crystal symmetry directions and
hence their diffraction. In general, rotating away from the symmetric configuration,
there will be fewer small reciprocal lattice vectors that can contribute to
diffraction, decreasing their influence. However, such a rotation cannot eliminate
the contributions from reciprocal lattice vectors of the type $\langle 0 \pm \!\!2\;
0 \rangle$, which are perpendicular to the incoming and outgoing beams.

\begin{vchfigure}
\includegraphics[width=0.85\textwidth,keepaspectratio=true,clip]
{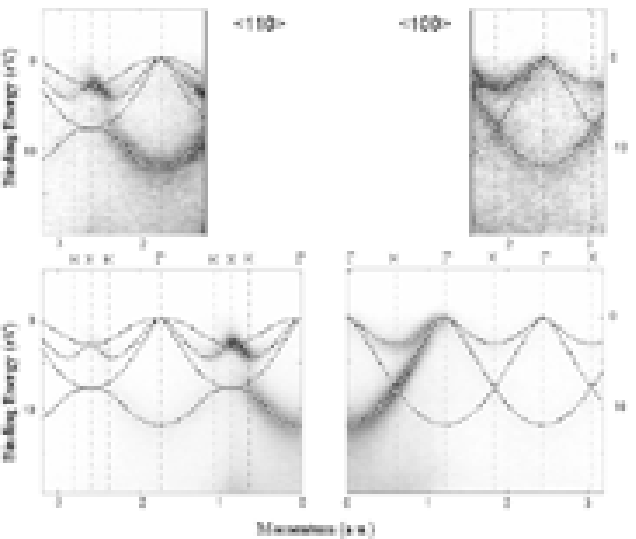} \vchcaption{Measured momentum densities along the $\langle 1  1 0
\rangle$ (left) and $\langle 1 0 0 \rangle$ (right) directions in linear grey-scale
with the LMTO band structure in the repeated zone scheme superimposed. The top
panels show the high momentum parts with an enhanced grey-scale to highlight the low
intensity diffracted contributions.} \label{emsfig8}
\end{vchfigure}
In Fig. \ref{emsfig8} we show the measured energy resolved momentum densities for
two high symmetry directions in a grey-scale plot with the LMTO band structure in
the repeated zone scheme superimposed on it. The contributions due to diffracted
beams can be seen  particularly at large momentum in the top panels with their
enhanced grey-scale. At ${\bm q} = 0$ the two spectral densities should be
identical, however, as can be seen from the figure, at this $\Gamma$ point there is
a small peak in the density close to zero binding energy in the $\langle 1 0 0
\rangle$ direction, which is absent in the $\langle 1 1 0 \rangle$ direction. This
density is due to the  $\langle 1 1 1 \rangle$ reciprocal lattice vector (the
shortest reciprocal lattice vector for the diamond lattice, length 1.06 a.u.)
shifting density from  $\Gamma_{(1,1,1)}$  to $\Gamma_{(1,0,0)}$. Rotation by
$10^\circ$ around the spectrometer $y-$axis removes this peak at low binding
energies \cite{Vos03B}. In this way, by examining the diffracted contributions in
regions where the spectral density should be zero, it is possible to remove the
diffracted components from the measured spectral momentum density. This is shown
explicitly in Fig. \ref{emsfig9}.

\begin{vchfigure}
\includegraphics[width=0.85\textwidth,keepaspectratio=true,clip]
{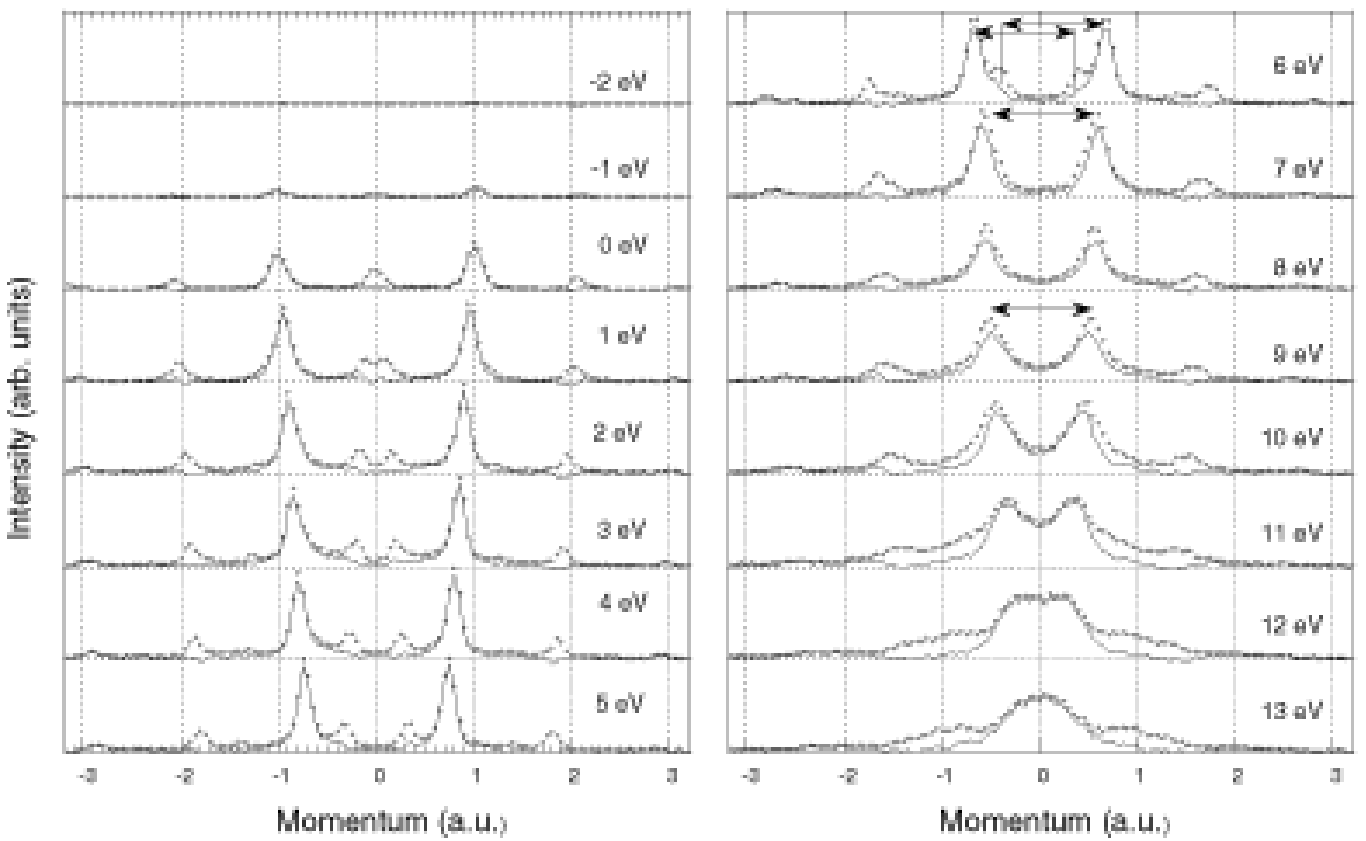} \vchcaption{The measured intensities (crosses) along $\langle 1 1 1
\rangle$ direction. The solid lines give the momentum densities corrected for
diffraction as discussed in the text. In the right panel the length of the
reciprocal lattice vector ${\bm G}=\langle 1 1 1 \rangle$ is indicated as well. }
\label{emsfig9}
\end{vchfigure}

Diffraction effects are most obvious when the $\langle 1 1 1 \rangle$ direction is
aligned with the spectrometer $y-$axis as can be seen  already in the bottom panel
of Fig. \ref{emsfig5}. The measured intensity distribution is given by the crosses
in Fig. \ref{emsfig9}, which shows cuts through the data at the indicated binding
energies. At the valence band maximum there are five peaks visible, at $q = 0$, $\pm
1.05$ a.u., and $\pm 2.1$ a.u. The dominant peaks at ± 1.05 a.u. are due to momentum
density at the top of band 2 (see Fig. \ref{emsfig3}), the other peaks are due to
diffraction by a crystal reciprocal lattice vector. By assuming that the diffracted
components add incoherently  to the real spectral momentum density, the latter can
be obtained by subtraction. The resulting momentum density is indicated by the solid
line in Fig. \ref{emsfig9}. The small gap between bands 1 and 2 (see Fig.
\ref{emsfig3}) shows up as a reduction in density around 8 eV.

After subtraction of the diffracted component the peak at the top of the valence
band at ${\bm q}=0$ disappears completely, indicating that it was entirely due to
diffraction with a reciprocal lattice vector of either ${\bm G}=\langle 1 1 1
\rangle$ or ${\bm G}=\langle -1 -1 -1 \rangle$.  However, in the binding energy
range near 6 eV there are clear shoulders left after subtraction.  This is to be
expected as near the Brillouin zone boundary these plane waves are expected to
contribute significantly to the Bloch waves. The shoulder and main line that are
part of the same Bloch function are shown in this figure, connected by a line of the
length of the reciprocal lattice vector.

\subsection{Many-body Effects}

Up to this point we have concentrated on those aspects of the electronic structure
of silicon that could be understood in the independent particle picture. However,
even from the semi-quantitative grey-scale presentations in Fig. \ref{emsfig5} (and
Fig. \ref{emsfig8}) of the measured intensities it is clear that they do not follow
the LMTO calculations in detail. The measurements show maximum intensity at
intermediate energies or near the top of the band, whereas the theory predicts
maximum intensity near the bottom of the band. Also the energy-widths of the
measured density distributions at a given momentum are much larger than the
theoretical ones even though the experimental energy resolution has been included in
the calculations. This is due to lifetime broadening of the spectral momentum
density by electron correlation effects. We will now look at some binding energy
spectra at appropriate momenta in more detail in order to explore the role of
many-body effects in the electronic structure of silicon.
\begin{vchfigure}
\includegraphics[width=0.92\textwidth,keepaspectratio=true,clip]
{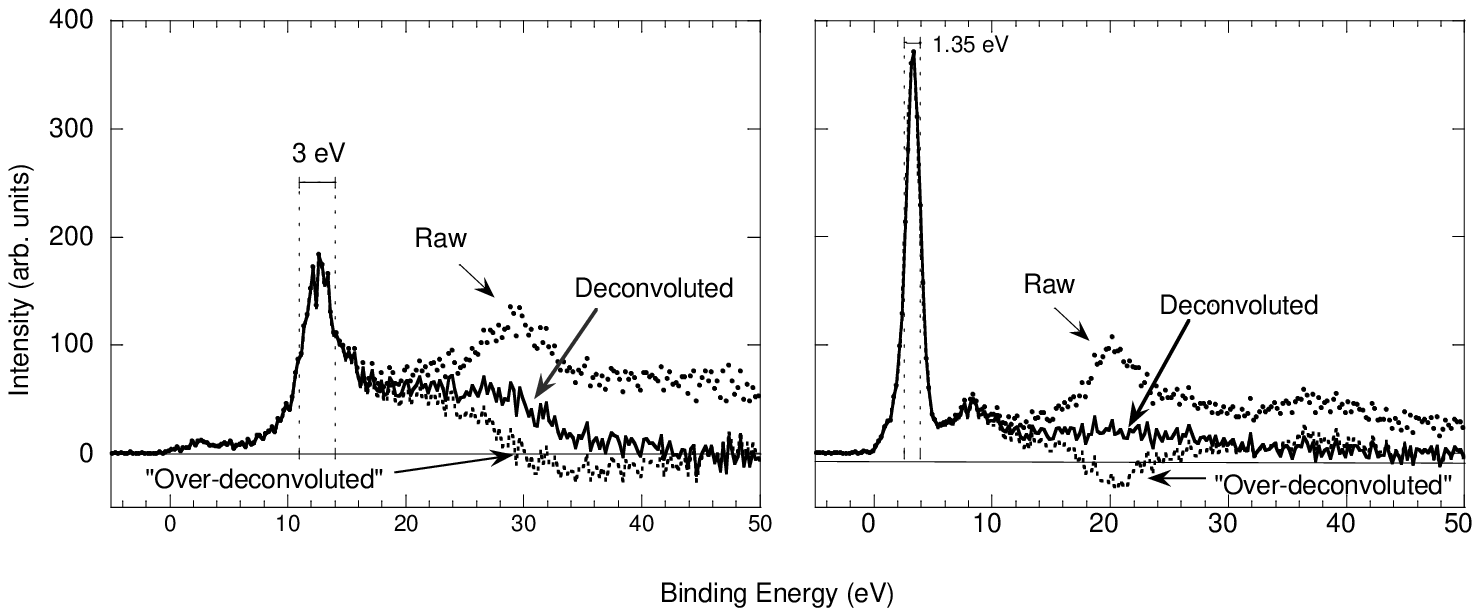} \vchcaption{EMS spectra  along the $\langle 1 1 0 \rangle$ direction
at $q_y\simeq 0$ near the $\Gamma$ point (left) and $q_y \simeq 0.87$ a.u. near the
X point (right). An identical deconvolution procedure is followed for both spectra.
The raw data are indicated by the dots, the data deconvoluted for inelastic
multiple-scattering contributions by the solid line, and the squares show data that
have deliberately been `over-deconvoluted' (see text for details). }
\label{emsfig10}
\end{vchfigure}

Fig. \ref{emsfig10} shows two spectra obtained along the $\langle 1 1 0 \rangle$
direction, one at $q \simeq 0$ near the $\Gamma$ point and the other at $q \simeq
0.87$ a.u. near the $X$ point. The peak near the $X$ point is very much narrower and
taller than the peak near $q \simeq 0$. The latter is also quite asymmetric. As well
as the raw data the figure includes a curve showing the data deconvoluted, in a
parameter-free way, for inelastic scattering using only the measured energy loss
spectrum (see Fig.\ref{emsfig2}) \cite{Vos02D}. This still leaves considerable
intensity at high binding energies, extending some 20 eV or more above the position
of the quasiparticle peak. If one tries to remove all the intensity at binding
energies above the quasiparticle peak by further deconvolution one obtains
non-physical negative intensities at some binding energies.

\begin{vchfigure}
\includegraphics[width=0.92\textwidth,keepaspectratio=true,clip]
{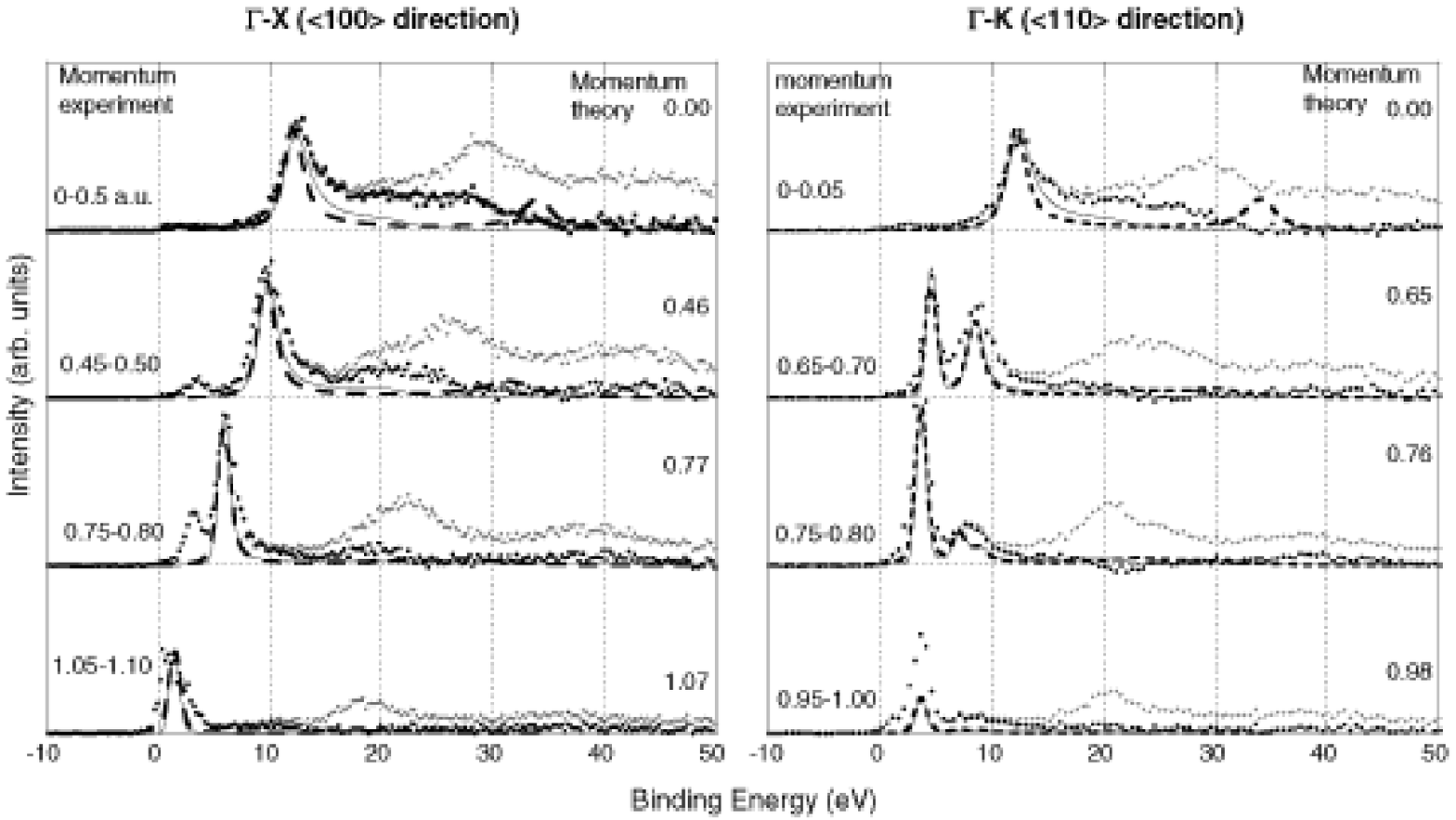} \vchcaption{EMS spectra, raw and deconvoluted for inelastic
multiple-scattering effects, at selected momenta along the $\langle 1 0 0 \rangle$
(top) and $\langle 1 1 0 \rangle$ (bottom) directions. The full and dashed lines are
the results of the cumulant expansion and $GW$ calculations, respectively, and they
are normalized to the quasiparticle peak at ${\bm q} = 0$.} \label{emsfig11}
\end{vchfigure}

In order to describe the data more quantitatively we performed full-scale many-body
calculations using the $GW$ and Cumulant Expansion approximations to the one-hole
Green's function (see section \ref{el_cor}). The results are presented in Fig
\ref{emsfig11}, which shows raw as well as the deconvoluted spectra along each of
the two symmetry directions compared with the first-principles many-body
calculations. The theories, convoluted with the experimental energy resolution, are
normalized to the data at one point only, namely to the peak of the quasiparticle
structure in the common ${\bm q} \simeq 0$ spectra, the theories having the same
total density in these spectra. The small low binding energy peak present in the
data in the $\langle 1 0 0 \rangle$ direction at intermediate $q_y$ values and not
in the calculations is due to the picking up of intensity from band 4 as discussed
earlier.

The \emph{GW} calculation gives a peak in the satellite intensity at around 34 eV at
${\bf q}\simeq 0$.  This is not observed in the data. This is a well-known failure
of the GW approximation \cite{Hedin99}.  The cumulant expansion calculation gives a
better fit to the data, although it too gives smaller satellite density than the
experiment. Both calculations describe the main quasiparticle features quite well in
all the spectra. In particular they give the broadening and the asymmetric structure
of this feature reasonably accurately at the smaller momenta. It is this broadening
which gives rise to the reduction in the peak heights at the lower momenta. However,
even though the calculations give significant life-time broadening, they
nevertheless underestimate the width of the quasiparticle structure, the cumulant
expansion model giving a slightly better fit.

\begin{vchfigure}
\includegraphics[width=0.5\textwidth,keepaspectratio=true,clip]
{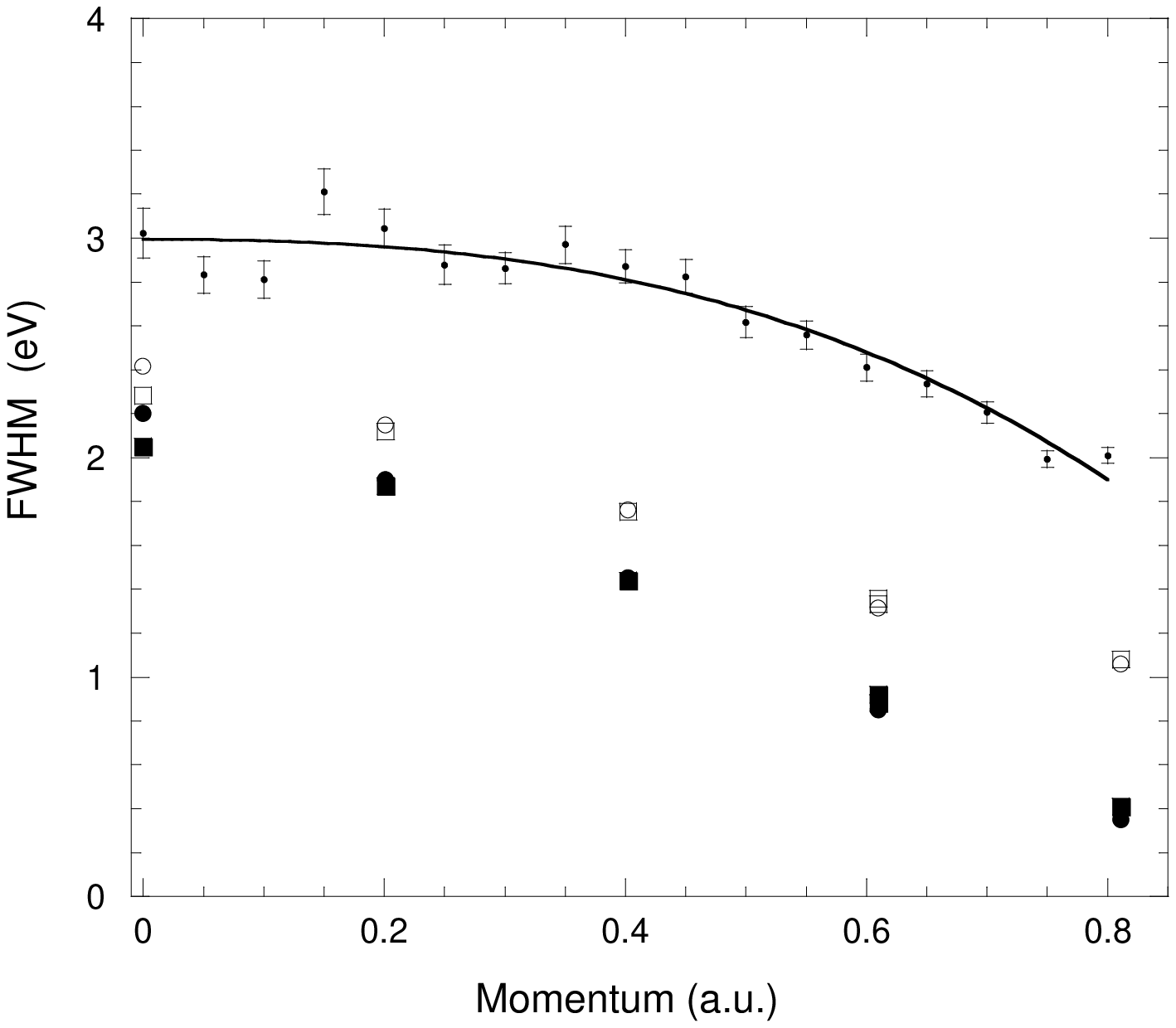} \vchcaption{The FWHM of the quasiparticle peaks in the $\langle 1 0 0
\rangle$ direction plotted as a function of momentum. The measurements are indicated
by the points with error bars. The thin line is a smooth fit to the experimentally
obtained FWHM.  The theoretical estimate of the peak width is indicated by squares
(\emph{GW} theory) and circles  (cumulant expansion theory). The filled symbols
refer to the calculated line width whereas the open symbols include the effect of
finite experimental energy resolution.} \label{emsfig12}
\end{vchfigure}
The finite quasiparticle (hole) lifetimes  causes broadening of the observed
features. In Fig. \ref{emsfig12} we plot as a function of the momentum  the observed
width (FWHM) for spectra obtained along the $\langle 1 0 0 \rangle$ direction. Near
zero momentum the binding energy is a weak function of momentum, and the observed
width should be a good indication of the life-time broadening.  Away from zero
momentum the binding energy becomes a strong function of momentum, and additional
broadening is observed due to the finite momentum resolution of the spectrometer as
initial states with slightly different momenta (and hence binding energy) contribute
to the spectra. In spite of this a decrease in width is observed with increasing
momentum. The sharpest spectra are observed where the dispersion goes again through
an extremum (now minimum binding energy) i.e. near the $X$ point ($\langle 1 0 0
\rangle$ direction, see also fig. \ref{emsfig10}) and the $L$ point ($\langle 1 1 1
\rangle$ direction). Here a width of 1.35 eV is observed, a width clearly dominated
by the energy resolution of the spectrometer. Also shown in the figure are the FWHM
given by the many-body calculations.  At zero momentum the calculated width is
significantly smaller than the observed width (even if the calculations are
broadened with the spectrometer energy resolution). The calculated width decreases
more rapidly with momentum, which is, at least in part, due to the fact that we did
not incorporate the effects of finite energy and momentum resolution in the
calculations.

The density of the quasiparticle structure and the total density along the $\langle
1 0 0 \rangle$ direction are plotted in fig. \ref{emsfig13}. Also shown are the
quasi-particle densities given by the cumulant expansion calculations, which
somewhat underestimate the satellite density. For this figure the satellite density
was defined as the total density above the main quasi-particle feature, whereas in
the experiment we take the intensity extending 20 eV above
 the end of the quasi-particle peak. in this way we obtain for the experiment that
 the
percentage of the total density that is contained in the satellite structure
decreases as the momentum increases. For instance, along the $\langle 1 0 0 \rangle$
direction, the satellite structure has essentially disappeared by momentum values
near 1 a.u., whereas at ${\bf q}=0$ it accounts for around 40 \% of the total
density (Fig. \ref{emsfig13}). In the theory the satellite intensity remains fairly
constant (at 25 \% of the total density) but becomes more spread out over binding
energy with increasing momentum.
\begin{vchfigure}
\includegraphics[width=0.5\textwidth,keepaspectratio=true,clip]
{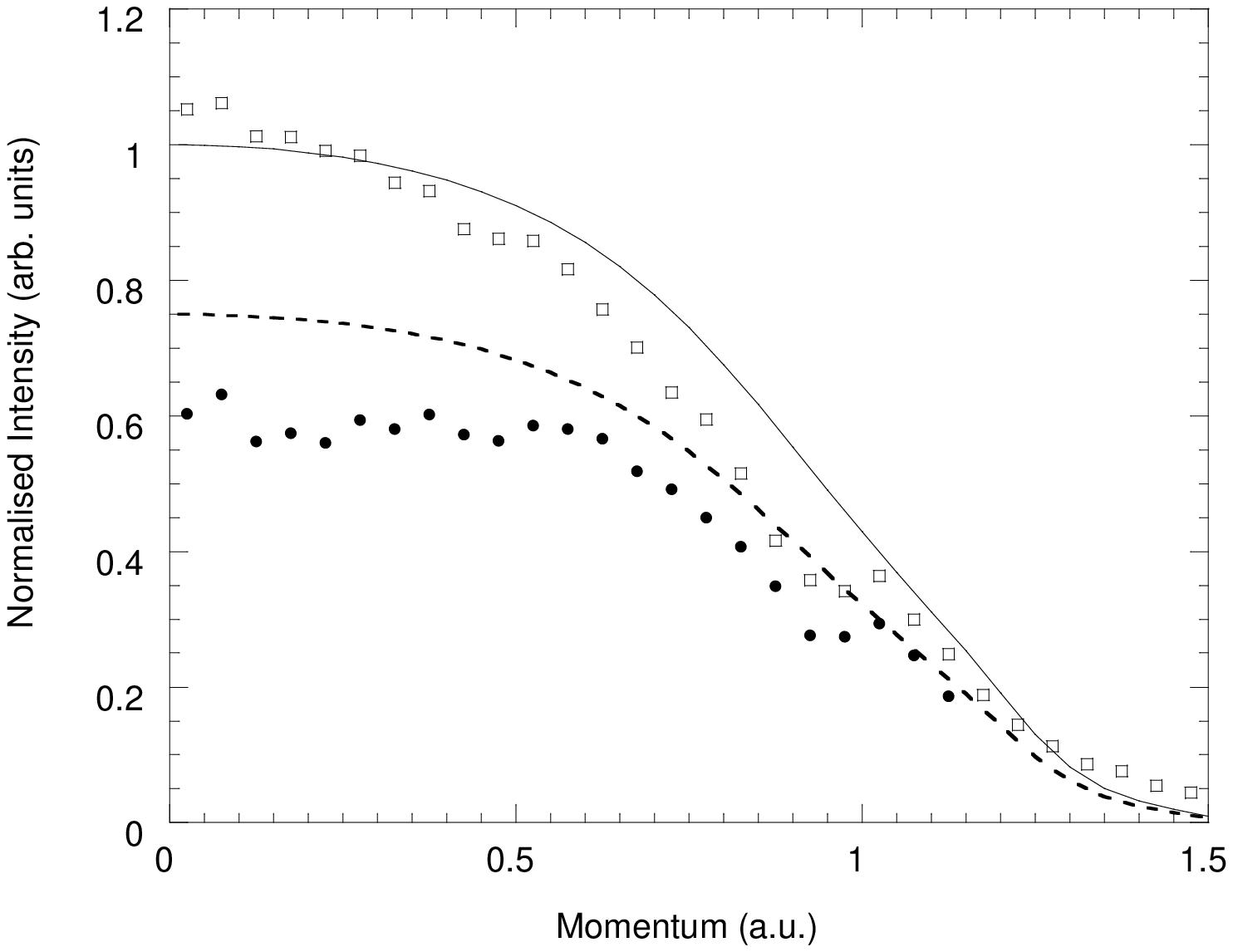} \vchcaption{The energy integrated electron momentum density along the
$\langle 1 0 0 \rangle$ direction. The open squares give the total density, and the
solid circles the quasi-particle density. Also included is the total density given
by the LMTO calculation (solid line) and the corresponding quasi-particle intensity,
as predicted by the cumulant expansion  calculations (dashed line).}
\label{emsfig13}
\end{vchfigure}
\section{Conclusions}
We have measured the spectral momentum density of the prototype semiconductor Si
along the three high symmetry directions, and several cuts between  the $\langle 1 0
0 \rangle$ and $\langle 1 1 0 \rangle$ directions, all going through zero momentum
(the $\Gamma_{(0,0,0)}$ point). In addition several measurements were made along the
$\langle 1 1 0 \rangle$ direction but with the cuts shifted from $\Gamma_{(0,0,0)}$
point to coincide with nearby Brillouin zone edges.

The band structure, i.e. the dispersion of the position of the peak in the
quasiparticle structure, is very well described by the FP-LMTO-DFT-LDA
approximation. The present EMS data are in better agreement with the theory than the
available ARPES data. The band-width was determined to be 11.85 $\pm$ 0.2 eV, in
good agreement with ARPES data (12.5 $\pm$0.6eV\cite{Grobman72}) and the LMTO theory
(11.93 eV).

Diffraction of the incident and/or the emitted electron beams can shift intensity
from the `direct' spectral density by reciprocal lattice vectors. Such effects are
observed but can often be distinguished from the `direct' intensity. Changing the
diffraction conditions for the incoming and/or outgoing electrons pinpoints which
part of the measured intensity is diffraction related. In this way contributions due
to diffraction can be removed on the assumption that there is no interference
between the direct and diffracted electron waves. The study of any possible
interference effects could lead to additional information on the structure of
silicon \cite{Vos03B}.

The spectra, especially at low momenta, show very large effects due to correlations.
In particular the quasiparticle structures are very broad due to the short hole
lifetimes, and much of the spectral density is found in higher energy satellite
structure. This satellite structure is quite smooth and extends to around 20 eV
above the quasiparticle peak. Although the satellite structure dominates the density
at low momentum, its density drops off as the momentum is increased. Similarly the
lifetime broadening diminishes as the momentum increases. Comparison with first
principles many-body calculations shows that the \emph{GW}  approximation predicts
the main quasiparticle features reasonably well, but cannot describe the satellite
structure. The cumulant expansion approximation also describes the quasiparticle
features qualitatively well, but does considerably better than the \emph{GW}  in its
description of the shape of the satellite density. Although it predicts the energy
distribution and the momentum dependence of the satellite structure quite well, it
nevertheless significantly underestimates the satellite intensity, particularly at
low momenta. The fact that agreement is better but still not perfect for the more
detailed model is encouraging in that a quantitative comparison between the present
EMS data and other representations of the many-body wave function should lead to new
levels of understanding.

\section{Acknowledgements}

We are grateful to the Australian Research Council for financial support.


\begin{thebibliography}{10}

\bibitem{Vos02D}
M.~Vos, A.~S. Kheifets, V.~A. Sashin, E.~Weigold, M.~Usuda, and
  F.~Aryasetiawan.
\newblock Quantitative measurement of the spectral function of aluminum and
  lithium by electron momentum spectroscopy.
\newblock {\em Phys. Rev. B}, 66:{155414}, 2002.

\bibitem{Weigold99}
E.~Weigold and I.~E. McCarthy.
\newblock {\em {E}lectron Momentum Spectroscopy}.
\newblock Kluwer Academic/Plenum, New York, 1999.

\bibitem{Vos00}
M.~Vos, G.~P. Cornish, and E.~Weigold.
\newblock A high-energy (e,2e) spectrometer for the study of the spectral
  momentum density of materials.
\newblock {\em Rev. Sci. Instrum.}, 71:3831--3840, 2000.

\bibitem{Vos00b}
M.~Vos and E.~Weigold.
\newblock Developments in the measurement of spectral momentum densities with
  (e,2e) spectrometers.
\newblock {\em J. Electron Spectrosc. Relat. Phenom.}, 112:93--106, 2000.

\bibitem{Fleszar97}
A.~Fleszar and W.~Hanke.
\newblock Spectral properties of quasiparticles in a semiconductor.
\newblock {\em Phys. Rev. B}, 56:10228--10232, 1997.

\bibitem{Kheifets95}
A.S. Kheifets and Y.Q. Cai.
\newblock Energy-resolved electron momentum densities of diamond-structure
  semiconductors.
\newblock {\em J. Phys.: Condens. Matter}, 7:1821--1833, 1995.

\bibitem{Borrmann87}
W.~Borrmann and P.~Fulde.
\newblock Exchange and correlation effects of the quasiparticle band structure
  of semiconductors.
\newblock {\em Phys. Rev. B}, 35:9569--9579, 1987.

\bibitem{Sturm92}
K.~Sturm, W.~Sch{\"u}lke, and J.~R. Schmitz.
\newblock Plasmon-{F}ano resonance inside the particle-hole excitation spectrum
  of simple metals and semiconductors.
\newblock {\em Phys. Rev. Lett}, 68:228--231, 1992.

\bibitem{Hybertsen86}
M.~S. Hybertsen and S.~G. Louie.
\newblock Electron correlation in semiconductors and insulators: Band gaps and
  quasiparticle energies.
\newblock {\em Phys. Rev. B}, 34:5390--5413, 1986.

\bibitem{Godby88}
R.~W. Godby, M.~Schl{\"u}ter, and L.~J. Sham.
\newblock Self-energy operators and exchange-correlation potentials in
  semiconductors.
\newblock {\em Phys. Rev. B}, 37:10159--10175, 1988.

\bibitem{Rohlfing93}
M.~Rohlfing, P.~Kr{\"u}ger, and J.~Pollmann.
\newblock Quasiparticle band-structure calculations for {C}, {Si}, {Ge},
  {GaAs}, and {SiC} using gaussian-orbital basis sets.
\newblock {\em Phys. Rev. B}, 48:17791--17805, 1993.

\bibitem{Engel96}
G.~E. Engel and W.~E. Pickett.
\newblock Investigation of density functionals to predict both ground-state
  properties and band structures.
\newblock {\em Phys. Rev. B}, 54:8420--8429, 1996.

\bibitem{Hedin99}
L.~Hedin.
\newblock On correlation effects in electron spectroscopies and the {GW}
  approximation.
\newblock {\em J. Phys: Condens. Matter}, 11:R489--R528, 1999.

\bibitem{Aryasetiawan98}
F.~Aryasetiawan and O.~Gunnarsson.
\newblock The {GW} method.
\newblock {\em Rep. Prog. Phys.}, 61:237, 1998.

\bibitem{Johansson90}
L.~S.~O. Johansson, P.~E.~S. Persson, U.~O. Karlsson, and R.~I.~G. Uhrberg.
\newblock Bulk electronic structure of silicon studied with angle-resolved
  photoemission from the {S}i(100) 2 x 1 surface.
\newblock {\em Phys. Rev. B}, 42:8991--8999, 1990.

\bibitem{Wachs85}
A.~L. Wachs, T.~Miller, T.~C. Hsieh, A.~P. Shapiro, and T.-C. Chiang.
\newblock Angle-resolved photoemission studies of {G}e(111)-c(2x8),
  {G}e(111)-{(1x1)H}, {S}i(111)-(7x7) and {S}i(100)-(2x1).
\newblock {\em Phys. Rev. B}, 32:2326--2333, 1985.

\bibitem{Rich89b}
D.~H. Rich, G.~E. Franklin, F.~M. Leibsle, T.~Miller, and T.~C. Chiang.
\newblock Synchrotron photoemission studies of the {S}b-passivated {S}i
  surfaces: Degenerate doping and bulk band dispersion.
\newblock {\em Phys. Rev. B}, 40:11804--11816, 1989.

\bibitem{Bansil02}
A.~Bansil, M.~Lindroos, S.~Sahrakorpi, R.~S. Markiewicz, G.~D. Gu, J.~Avila,
  L.~Roca, A.~Tejeda, and M.~C. Asensio.
\newblock First principles simulations of energy and polarization dependent
  angle-resolved photoemission spectra of {B}i2212.
\newblock {\em J. Phys. Chem. Solids}, 63:2175--2180, 2002.

\bibitem{Bell01}
F.~Bell and J.~R. Schneider.
\newblock Three-dimensional electron momentum densities of solids.
\newblock {\em J. Phys.: Condens. Matter}, 13:7905--7922, 2001.

\bibitem{Sattler01}
T.~Sattler, Th. Tschentscher, J.~R. Schneider, M.~Vos, A.~S. Kheifets, D.~R.
  Lun, E.~Weigold, G.~Dollinger, H.~Bross, and F.~Bell.
\newblock The anisotropy of the electron momentum density of graphite studied
  by ($\gamma$,e$\gamma$) and (e,2e) spectroscopy.
\newblock {\em Phys. Rev. B}, 63:155204, 2001.

\bibitem{Itou99}
M.~Itou, S.~Kishimoto, H.~Kawata, M.~Ozaki, H.~Sakurai, and F.~Itoh.
\newblock Three-dimensional electron momentum density of graphite by (x,ex)
  spectroscopy with a time of flight electron energy spectrometer.
\newblock {\em J. Phys. Soc. Japan}, 68:515, 1999.

\bibitem{Vos04}
M.~Vos, V.~A. Sashin, C.~Bowles, A.~S. Kheifets, and E.~Weigold.
\newblock Probing the spectral densities over the full three-dimensional
  momentum space.
\newblock {\em J. Phys. Chem. Solids}, in press, 2004.

\bibitem{Utteridge00}
S.~J. Utteridge, V.~A. Sashin, S.~A. Canney, M.~J. Ford, Z.~Fang, D.~R. Oliver,
  M.~Vos, and E.~Weigold.
\newblock Preparation of a 10 nm thick single-crystal silicon membrane
  self-supporting over a diameter of 1 mm.
\newblock {\em Appl. Surf. Sci.}, 162-163:359--367, 2000.

\bibitem{Vos96}
M.~Vos and M.~Bottema.
\newblock Monte {C}arlo simulations of (e,2e) experiments in solids.
\newblock {\em Phys. Rev. B}, 54:5946--5954, 1996.

\bibitem{Vos02c}
M.~Vos, A.~S. Kheifets, and E.~Weigold.
\newblock Electron momentum spectroscopy of metals.
\newblock In D.~H. Madison and M.~Schulz, editors, {\em Correlations,
  Polarization and Ionization in Atomic Systems, {IAP} Conference Proceedings
  604}, pages 70--75, New York, 2002. American Institute of Physics.

\bibitem{Jones89}
R.O. Jones and O.~Gunnarsson.
\newblock The density functional formalism, its applications and prospects.
\newblock {\em Rev. Mod. Phys.}, 61:689--746, 1989.

\bibitem{Dreizler90}
R.M. Dreizler and E.K.U. Gross.
\newblock {\em Density Functional Theory}.
\newblock Springer Verlag, Berlin, 1990.

\bibitem{Pickett89}
W.E. Pickett.
\newblock Electronic structure of the high-temperature oxide superconductors.
\newblock {\em Rev. Mod. Phys.}, 61:433--512, 1989.

\bibitem{Skriver84}
H.L. Skriver.
\newblock {\em The {LMTO} Method}.
\newblock Springer Verlag, Berlin, 1984.

\bibitem{Kheifets99}
A.~S. Kheifets, D.~R. Lun, and S.~Yu Savrasov.
\newblock Full-potential linear-muffin-tin-orbital calculation of electron
  momentum densities.
\newblock {\em J. Phys.: Condens. Matter}, 11:6779--6792, 1999.

\bibitem{Hedin65}
L.~Hedin.
\newblock Method for calculating the one-particle green's function with
  application to the electron-gas problem.
\newblock {\em Phys. Rev.}, 139:A796--A823, 1965.

\bibitem{Hedin69}
L.~Hedin and S.~Lundqvist.
\newblock {\em Solid State Physics}, 23:1--181, 1969.

\bibitem{Langreth70}
D.~C. Langreth.
\newblock Singularities in x-ray spectra of metals.
\newblock {\em Phys. Rev. B}, 1:471, 1970.

\bibitem{Bergersen73}
B.~Bergersen, F.~W. Klus, and C.~Blomberg.
\newblock Single particle {G}reen's function in the electron-plasmon
  approximation.
\newblock {\em Can.J.Phys.}, 51:102, 1973.

\bibitem{Hedin80}
L.~Hedin.
\newblock Effects of recoil on shake-up spectra in metals.
\newblock {\em Physica Scripta}, 21:477--480, 1980.

\bibitem{Aryasetiawan96}
F.~Aryasetiawan, L.~Hedin, and K.~Karlsson.
\newblock Multiple plasmon satellites in {Na} and {Al} spectral functions from
  ab initio cumulant expansion.
\newblock {\em Phys. Rev. Lett.}, 77:2268--2271, 1996.

\bibitem{Vos99}
M.~Vos, A.~S. Kheifets, E.~Weigold, S.~A. Canney, B.~Holm, F.~Aryasetiawan, and
  K.~Karlsson.
\newblock Determination of the energy-momentum densities of aluminium by
  electron momentum spectroscopy.
\newblock {\em J. Phys.: Condens. Matter}, 11:3645, 1999.

\bibitem{Vos01}
M.~Vos, A.~S. Kheifets, and E.~Weigold.
\newblock The spectral momentum density of aluminum measured by electron
  momentum spectroscopy.
\newblock {\em J. Phys. Chem. Solids}, 62:2215--2221, 2001.

\bibitem{Heine60}
V.~Heine.
\newblock {\em Group Theory in Quantum Mechanics}.
\newblock Pergamonn, New York, 1960.

\bibitem{Himpsel83}
F.J. Himpsel, G.~Hollinger, and R.A. Pollak.
\newblock Determination of the fermi-level pinning position at si(111)
  surfaces.
\newblock {\em Phys. Rev. B}, 28:7014--7018, 1983.

\bibitem{Vos03B}
M.~Vos, A.~S. Kheifets, V.~A. Sashin, and E.~Weigold.
\newblock Influence of electron diffraction on measured energy-resolved
  momentum densities in single-crystalline silicon.
\newblock {\em J. Phys Chem. Solids}, 64:2507--2515, 2003.

\bibitem{Allen90}
L.~J. Allen, I.~E. McCarthy, V.~W. Maslen, and C.~J. Rossouw.
\newblock Effects of diffraction on the (e,2e) reaction in crystals.
\newblock {\em Aust. J. Phys.}, 43:453--464, 1990.

\bibitem{Matthews93}
R.S. Matthews.
\newblock PhD thesis, Flinders University of South Australia, 1993.

\bibitem{Grobman72}
W.D. Grobman and D.E. Eastman.
\newblock Photoemission valence-band densities of states for si, ge, and {GaAs}
  using synchrotron radiation.
\newblock {\em Phys. Rev. Lett.}, 29:1508--1512, 1972.

\end{thebibliography}
\end{document}